\documentclass[%
 aip,
 jmp,%
 amsmath,amssymb,
preprint,%
]{revtex4-1}

\usepackage{graphicx}
\usepackage{dcolumn}
\usepackage{bm}
\usepackage{epstopdf}
\usepackage{amssymb}
\usepackage{color}
\usepackage[colorlinks=true, letterpaper=true, pdfstartview=FitV, linkcolor=blue, citecolor=blue, urlcolor=blue]{hyperref}
\usepackage{lineno}

\begin{document}

\title{Two-Dimensional Ferromagnetic Half-Metallic Janus V$_2$AsP Monolayer}
\author{Qiuyue Ma}
\author{Guochun Yang}
\author{Yong Liu}\email{yongliu@ysu.edu.cn}
\affiliation{State Key Laboratory of Metastable Materials Science \& Technology, Yanshan University, Qinhuangdao 066004, China }
\affiliation{Key Laboratory for Microstructural Material Physics of Hebei Province, School of Science, Yanshan University, Qinhuangdao 066004, China }
\begin{abstract}
Two-dimensional (2D) ferromagnetic materials present promising candidates for spintronic devices, and the half-metallic materials with 100\% spin polarization at Fermi energy level are highly desired for many spin-based devices. 2D Janus materials have attracted great attention in recent years due to their excellent properties induced by breaking the symmetry. Here, using the density functional theory, we report that the Janus V$_2$AsP monolayer demonstrates a charming ferromagnetic half-metallic feature. It is dynamically stable in view of the absence of imaginary frequency phonon. The half-metallic gap is about 0.38 eV and the spin splitting of about 1.34 eV for the V$_2$AsP monolayer. Interestingly, a tensile strain of 4.9\% can induce it to undergo a phase transition from ferromagnetic to anti-ferromagnetic state. Moreover, the Curie temperature (\emph{T}$_c$) enhances with the increase of compressive strain. All these appealing properties make the half-metallic Janus V$_2$AsP monolayer a promising material for 2D spintronic applications.
\end{abstract}

\maketitle


\maketitle
\section{INTRODUCTION}
Spintronics, which uses the spin degrees of freedom of electrons for information transmission, storage and processing, has attracted extensive attention because of its unique advantages of low power consumption, fast data processing speed and high integration density~\cite{1S. A. Wolf-Science-2001}. Two-dimensional (2D) ferromagnets have potential applications in nanoscale spintronic devices. A large number of 2D materials~\cite{2K. S. Novoselov-Science-2004,3K. S. Novoselov-Proceedings-2005,4C. Jin-Phys. Rev. Lett.-2009,5M. Chhowalla-Nature Chemistry-2013,6K. F. Mak-Phys. Rev. Lett.-2010,7A. Splendiani-Nano Lett.-2010} have been discovered in recent years. However, the lack of intrinsic ferromagnetism heavily restricts their application in spintronic devices. The ferromagnetic (FM) ordering of 2D materials can be obtained by doping~\cite{8B. Wang-Nanoscale Horiz.-2018,9N. Miao-J. Am. Chem. Soc.-2017,10B. Li-Nature Communications-2017}, external strain~\cite{11B. Xu-Applied Physics Letters-2020,12Y. Ma-ACS Nano-2012}, external electric field~\cite{13Z. Fei-Nature Materials-2018} and defect engineering~\cite{14Z. Zhang-ACS Nano-2013,15J. Guan-ChemPhysChem-2013,16Y. Tong-Advanced Materials-2017}. Where external strain is an effective approach to adjust the electronic structures and magnetic properties of the low-dimensional materials. Many theoretical reports have indicated the tunable electronic structures and magnetic characteristics in strained monolayers~\cite{Y. Ma-ACS Nano-2012,S. Bertolazzi-ACS Nano-2011,X. Chen-Phys. Lett. A-2015,Y. Ma-Nanoscale-2011,L. Kou-ACS Nano-2011}. In the experiment, the application of adjustable biaxial strain to 2D materials has made remarkable progress~\cite{F. Ding-Nano Lett.-2010}. Additionally, under a biaxial tensile strain of approximately 13\% in MnPSe$_3$~\cite{17Q. Pei-Frontiers of Physics-2018}, it occurs a magnetic phase from antiferromagnetic (AFM) to FM state, and this transition is also achieved by electron and hole doping induce~\cite{18X. Li-J. Am. Chem. Soc.-2014}. The tunable magnetic properties of 2D ferromagnets have attracted great interest. On the other hand, the discovery of 2D magnetic materials with high spin polarization, large magnetic anisotropy energy (MAE), and high Curie temperature (\emph{T}$_c$) would promote the development of spintronic devices, and also would provide new opportunities in low-power-consumption spintronics and quantum computing, among many other applications~\cite{Z. Fei-Nat. Mater-2018,Y. Deng-Nature-2018,Y. Deng-Nature-2018}. Half-metallic materials are conducting in one spin channel but insulating in another optional channel which exhibiting 100\% spin polarization~\cite{19X. Li-J. Am. Chem. Soc.-2014,20M. Frik-New Journal of Physics-2007}. These properties of half-metallic materials are highly desired in many spin-based devices. Therefore, the 2D half-metals are ideal materials for spintronic nanoscale devices~\cite{21C. Felser-Angewandte Chemie International Edition-2007}.

In fact, the electronic structures largely determine the physical properties of materials, due to the destruction of structural symmetry of low dimensional materials, it can be significantly modulated. In recent years, 2D Janus materials have received extensive attention in the research field. The Janus monolayers due to out-of-plane asymmetry exhibit extraordinary physical characteristics, such as the piezoelectric polarization~\cite{22C. Zhang-Nano Lett.-2019} and preferred catalytic performance~\cite{23D. Er-Nano Lett.-2018,24X. Ma-J. Mater. Chem. A-2018}. The 2D Cr$_2$XS$_3$ (X = Br, I) are room-temperature magnetism Janus semiconductors by substituted one layer of halogon atoms with sulfur atoms to break symmetry of CrX$_3$ (X = Br, I) monolayers~\cite{25D. Wu-J. Phys. Chem. Lett.-2021}. Janus transition metal dichalcogenides MXY (M = Mo, W; X, Y = S, Se, Te; X $\neq$ Y)~\cite{26L. Dong-ACS Nano-2017} and Janus MoSSe monolayer show intrinsic dipole and piezoelectric effects~\cite{27A.-Y. Lu-Nature Nanotechnology-2017}. Multi-Functional Janus vanadium dichalcogenides VXX' (X/X' = S, Se, Te) also exhibit excellent physical properties, and their Curie temperature can be enhanced by the built-in electric field effect~\cite{S. Ji-Chin. Phys. Lett-2020}. Other Janus materials also show excellent properties, such as FeXY  (X, Y = Cl, Br, and I, X $\neq$ Y)~\cite{28R. Li-Nature Nanotechnology-2017}, V$_2$X$_3$Y$_3$(X, Y = Cl, Br and I; X $\neq$ Y)~\cite{29Y. Ren-Phys. Rev. B-2020},M$_2$SeTe(M = Ga, In)~\cite{30Y. Guo-Appl. Phys. Lett.-2017}, and the single-sided hydrogenated graphene~\cite{31J. Zhou-Nano Lett.-2009}. Overall, 2D Janus materials greatly promote the development of spintronic devices and expect to have potential applications in electronic and electromechanical devices.

In this work, based on first-principles, we systematically investigated the electronic and magnetic properties of Janus V$_2$AsP monolayer, demonstrating highly mechanical and dynamic stability. The Janus V$_2$AsP monolayer with large half-metallic band gap and spin band gap is an FM half metallic material. Moreover, we applied a biaxial strain to Janus V$_2$AsP monolayer. It is accompanied by the transition from an FM to AFM. Furthermore, the predicted \emph{T}$_c$ of V$_2$AsP monolayer is 83 K by Monte Carlo simulations. And the \emph{T}$_c$ can be enhanced with the increase of compressive strain.

\section{Methods}

The calculations of this work were performed by using the Vienna ab initio simulation software package (VASP)~\cite{32G. Kresse-Phys. Rev. B-1993,33G. Kresse-Phys. Rev. B-1993} based on the spin polarization density function theory (DFT)~\cite{34W. Kohn-Phys. Rev.-1965,35P. Hohenberg-Phys. Rev.-1964}. The exchange correlation potential was calculated under the Perdew-Burke-Ernzerhof (PBE) function of Generalized Gradient Approximation (GGA)~\cite{36J. P. Perdew-Phys. Rev. Lett.-1996}. The plane-wave cutoff energy was set to be 500 eV. The tolerance criterion of energy and force were set to be 10$^{-6}$ and 0.01 eV/{\AA}, respectively. A $9\times9\times1$ Monkhorst-Pack special k-point mesh was used in Brillouin zone~\cite{37H. J. Monkhorst-Phys. Rev. B-1976}. A vacuum slab of 20 {\AA} was added along the z axis to avoid the interactions between the adjacent monolayers. The strong field Coulomb interaction was considered by using the GGA+U method, onsite Coulomb interaction parameter was set to be 3 eV based on the relevant previous reports~\cite{38J. He-J. Mater. Chem. C-2016}. Phonon dispersion spectrum was  analyzed based on density functional perturbation theory (DFPT) as implemented in the Phonopy code~\cite{39A. Togo-Phys. Rev. B-2008}.

\begin{figure}[t!hp]
\centerline{\includegraphics[width=0.85\textwidth]{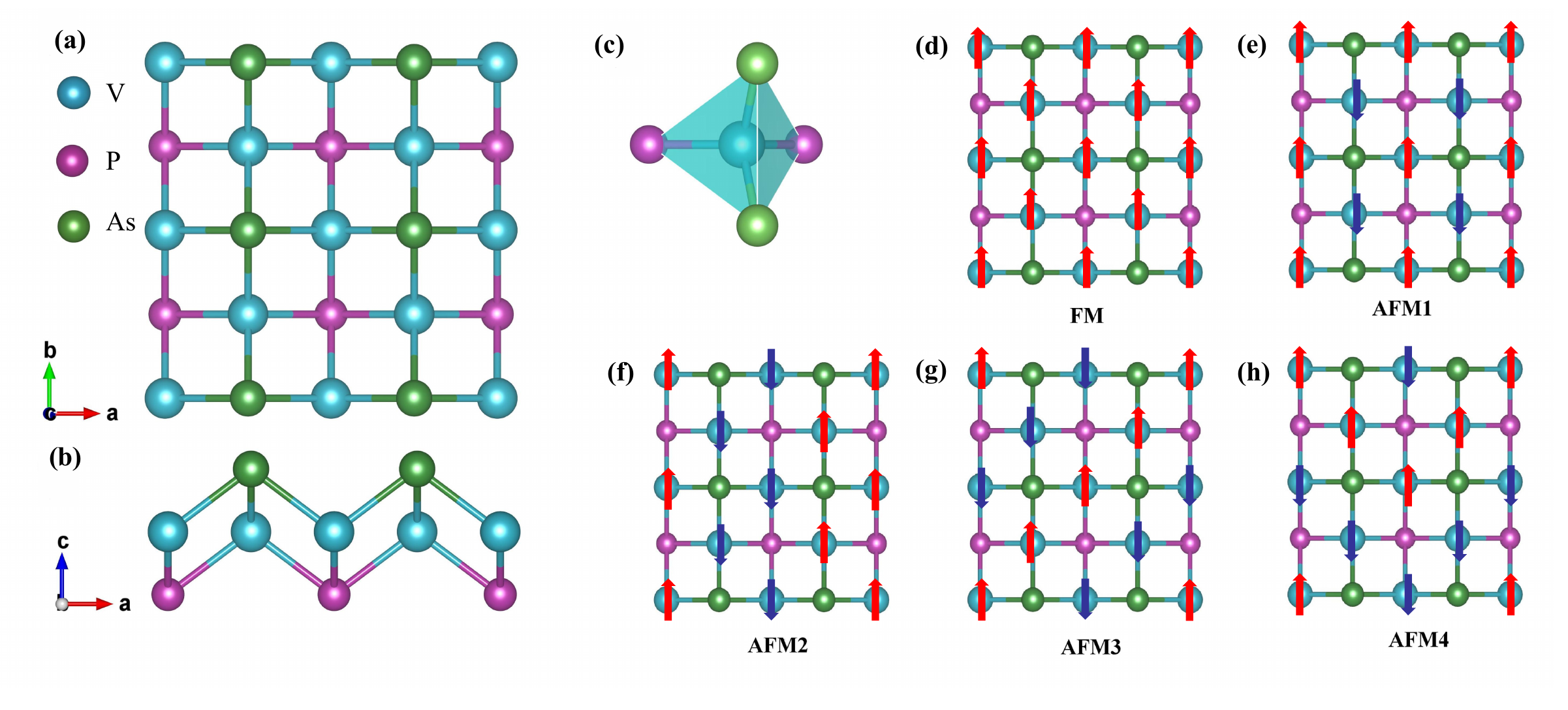}}
\caption{(a) Top and (b)(c) side views of Janus V$_2$AsP monolayer. (d)-(h) One Ferromagnetic (FM) and four antiferromagnetic (AFM) configurations.
\label{fig:stru}}
\end{figure}

\begin{figure}[tbp!]
\centerline{\includegraphics[width=0.80\textwidth]{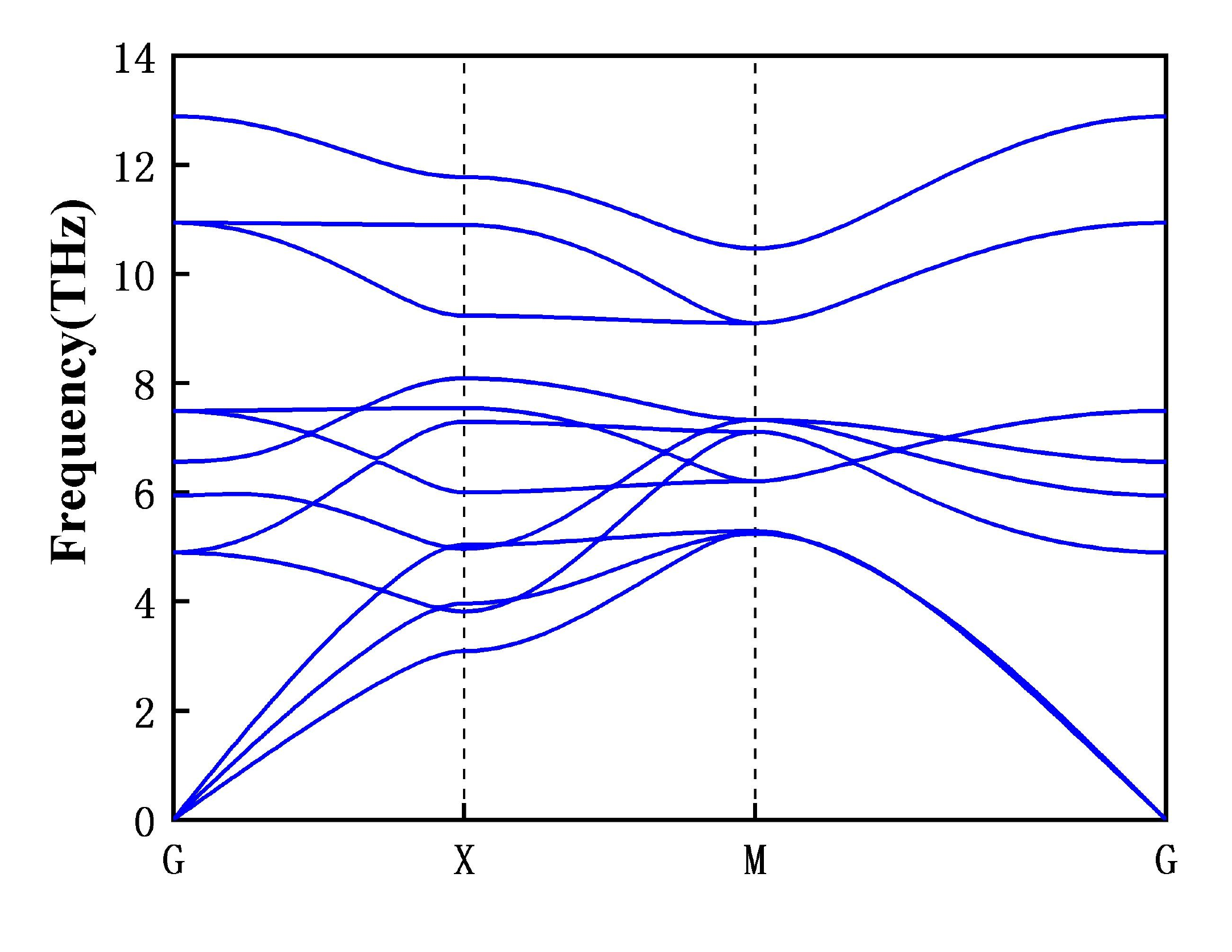}}
\caption{Phonon spectrum of the Janus V$_2$AsP monolayer
\label{fig:64-E-m}}
\end{figure}

\begin{figure}[tbp!]
\centerline{\includegraphics[width=0.95\textwidth]{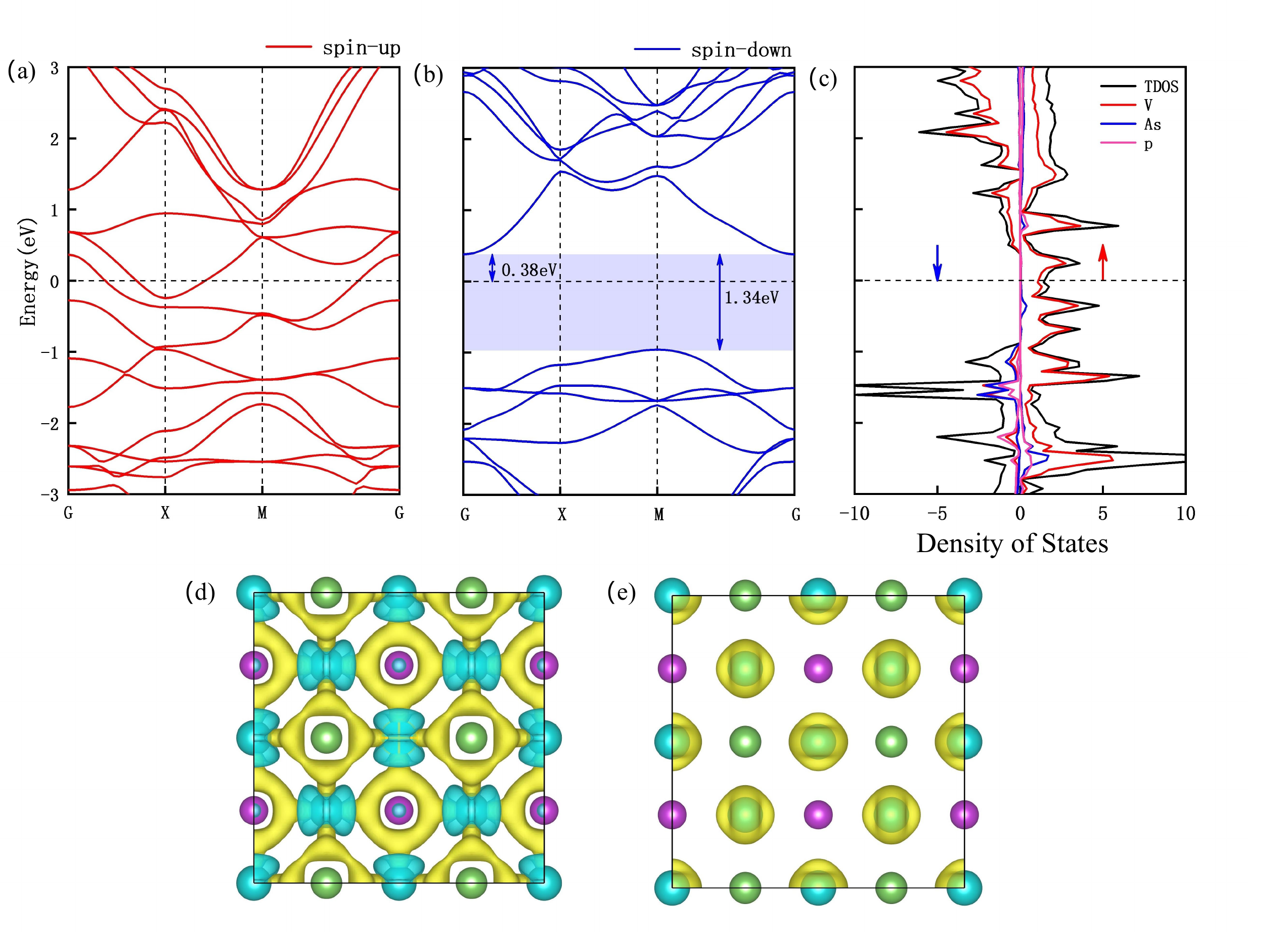}}
\caption{(a)(b) The band structures, (c)total density of states (TDOS) and projected density of states (PDOS) for Janus V$_2$AsP monolayer; (d)charge density difference isosurface plot and (e)spin density isosurface plots for ground state spin arrangements
\label{fig:64-E-k}}
\end{figure}

\section{ RESULTS and DISCUSSION}
As shown in Fig.~\ref{fig:stru}(a)-(c), the Janus V$_2$AsP monolayer consists of one V layer sandwiched  between the As layer and P layer with lattice parameters a = b = 4.43 {\AA}, which is isostructural to the MnX (X=P, As) monolayer~\cite{B. Wang-Nanoscale-2019}. The bond lengths of V-As, V-P, and V-V are 2.54 {\AA}, 2.40 {\AA}, and 3.13 {\AA}. The $2\times2\times1$ supercell with one FM and four AFM configurations shown in Fig.~\ref{fig:stru}(d)-(h) are considered to ascertain the magnetic ground state of Janus V$_2$AsP monolayer. We calculated the energy difference ($\Delta$E = E$_F$$_M$ $-$ E$_A$$_F$$_M$) relative to FM configurations are 33.32, 21.18, 45.98, and 76.94 meV per V atom for AFM1, AFM2, AFM3, and AFM4 configurations, our results suggest that the magnetic ground state of Janus V$_2$AsP monolayer is FM. Besides, the non-magnetic (NM) state can be neglected due to the tremendous energy difference between NM state and magnetic states.

Next, the elastic stiffness tensors C$_{11}$ (C$_{11}$ = C$_{22}$) and C$_{12}$ were calculated as 28.95 and 8.90 N/m. It indicates that the Janus V$_2$AsP monolayer is mechanically stable according to the Born stability criterion (C$_{11}{>}$0, C$_{22}{>}$0 and C$_{11}$-C$_{12}{>}$0)~\cite{40Z.-j-Phys. Rev. B-2007}. The Young's modulus (\emph{Y}$_{2D}$) and Poission's ratio were evaluated to be 26.21 N/m and 0.31, respectively. The value of Young's modulus is lower than that of MoS$_2$~\cite{41S. Bertolazzi-ACS Nano-2011}, indicating that V$_2$AsP monolayer is softer and can sustain a large strain. To verify the dynamic stability of V$_2$AsP monolayer, phonon spectrum was calculated. The phonon dispersion relation has no imaginary frequency (Fig.~\ref{fig:64-E-m}), indicating that Janus V$_2$AsP monolayer is dynamically stable.

The band structures and density of states for V$_2$AsP monolayer are shown in Fig.~\ref{fig:64-E-k}(a)(b). We find that the spin-up channels cross the Fermi  energy level, while the spin-down channels have band gap. Therefore, the Janus V$_2$AsP monolayer is a half-metallic material which satisfies 100\% spin polarization. Half-metallic materials are semiconductors in one spin channel and in another optional channel exhibit metallic character, which have potential applications for spintronic devices~\cite{42C. Felser-Angewandte-2007}. The half-metallic band gap and spin gap with PBE+U functional are 0.38 eV and 1.34 eV. The PBE+U method always underestimates the band gap, then we used the Heyd-Scuseria-Ernzerh (HSE06) hybrid functional method to obtain more accurate electronic structures(in Fig.S1, supplementary material). Note that the half-metallic gap (0.61 eV) and the band gap of spin-down channel (1.93 eV) are larger than that of PBE+U method. Fig.~\ref{fig:64-E-k}(c), the density of states show that the spin polarization at the Fermi level mainly derives from the V atoms, the valence band maximum (VBM) of the spin-down channel is mainly contributed by V, As and P atoms, while the contribution of conduction band minimum (CBM) is V atoms. Fig.~\ref{fig:64-E-k}(d) shows the differential charge density which is the difference between the charge density at the bonding point and the atomic charge density at the corresponding point of Janus V$_2$AsP monolayer. The yellow and blue regions represent the net charge gain and loss, respectively. It is obvious that As and P atoms gain electrons, while the V atoms lose electrons. The V-As and V-P bonds show more ionic. As illustrated in Fig.~\ref{fig:64-E-k}(e), the local magnetic moments are mainly the contribution of V atom which is consistent with previous analysis of magnetic moment.

To estimate the \emph{T}$_c$ of Janus V$_2$AsP monolayer, we performed Monte Carlo simulations of the Heisenberg model. The Hamiltonian:

\begin{eqnarray}
{H =  - \sum\limits_{ < ij > } {{J_{1}}{S_i}{S_j}}- \sum\limits_{ < ik > } {{J_{2}}{S_i}{S_k}}- A{{S_i^Z}{S_i^Z}}},
\end{eqnarray}

where J$_1$ and J$_2$ are the exchange parameters between the nearest and the next-nearest, S is the spin vector of V atom,  S$_i^Z$ is the spin along z direction, and A is the parameter of magnetic anisotropy. To extract the exchange parameters, we utilize:

\begin{eqnarray}
{E(FM) = {E_0} - 2{J_1}{S^2}- 2{J_2}{S^2}- A{S^2}},
\end{eqnarray}

\begin{eqnarray}
{E(AFM1) = {E_0} + 2{J_1}{S^2}- 2{J_2}{S^2}- A{S^2}},
\end{eqnarray}

\begin{eqnarray}
{E(AFM3) = {E_0} + 2{J_2}{S^2}- A{S^2}},
\end{eqnarray}

Where E$_0$ is the energy without spin polarization, and the exchange parameters J$_1$ and J$_2$ of V$_2$AsP monolayer are calculated to be 8.3 and 7.3 meV per V atom. The result is shown in Fig.S2, supplementary material; indicating the \emph{T}$_c$ of Janus V$_2$AsP monolayer is 83 K. It is significantly that the \emph{T}$_c$ is higher than CrI$_3$ monolayer (45 K)~\cite{43B. Huang-Nature-2017}, Cr$_2$Ge$_2$Te$_6$ bilayer (30 K)~\cite{44C. Gong-Nature-2017}, CrSiTe$_3$ (35.7 K) and CrGeTe$_3$ (57.2 K)~\cite{45C. Gong-Nature-2017}. The determination of easy axis is able to study the magnetic coupling effect. In Fig.~\ref{fig:64-E-b}(a), we plot $\Delta$E = E$_a${$_b$  $-$ E$_{\theta}$ as a function of $\theta$, the E$_a${$_b$} and E$_{\theta}$ represent the energy difference between the spin directions of parallel and $\theta$ angled to the ab plane. We can find that the V$_2$AsP monolayer prefers an in-plane array and tend to decrease. Magnetic anisotropy is the key to the establishment of long-range 2D ferromagnetism. MAE mainly comes from the influence of spin-orbit coupling (SOC), and it can effectively counteract the influence of thermal fluctuation. MAE is defined as the energy difference between in plane and out of plane magnetization directions, the MAE of Janus V$_2$AsP monolayer is calculated as 177.39 $\mu$eV, which easy axis is in-plane spin arrays. The observed large MAE indicates that V$_2$AsP monolayer has potential for application in magnetic storage devices.

Besides, we analyzed the effect of biaxial strain on the properties of V$_2$AsP monolayer. The strain $\varepsilon$ is defined as (a $-$ a$_0$)$/$a$_0$, where a$_0$ and a are the lattice constant for the unstrained and strained system, respectively. As shown in Fig.~\ref{fig:64-E-b}(b), the red line represents the energy difference ($\Delta$E) between FM and AFM configurations, illustrating that the $\Delta$E increases with the decrease in compressive strain and the increase in tensile strain. when the $\Delta$E becomes greater than zero, the phase transition from FM to AFM state occurs at 4.9\% tensile strain. Meanwhile, the black line shows the change of Curie temperature under biaxial strain. With the decrease of $\varepsilon$, the Curie temperature tends to increase. We estimated that a compressive strain of -5\% can make the \emph{T}$_c$ reach to 130 K.

\begin{figure}[tbp!]
\centerline{\includegraphics[width=0.95\textwidth]{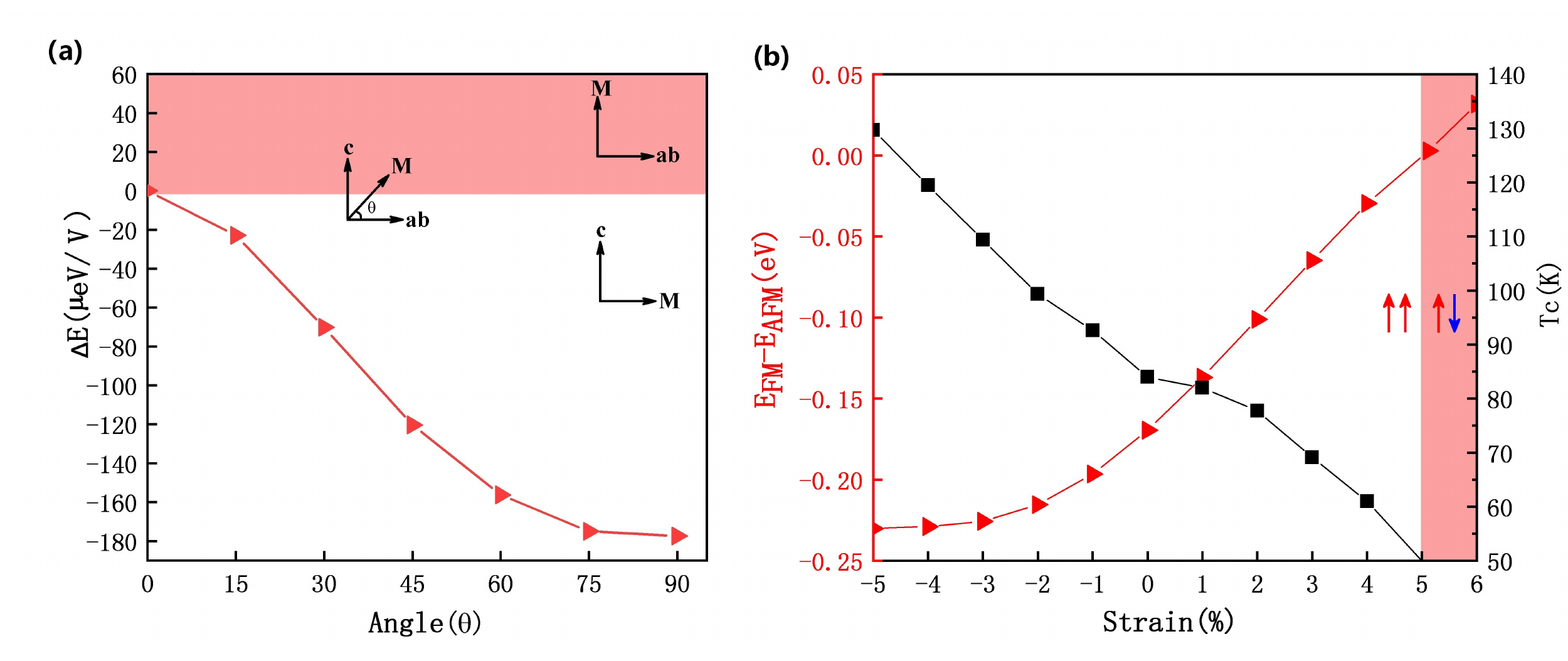}}
\caption{(a) Energy difference $\Delta$E between E$_a${$_b$} and E$_{\theta}$ per unit cell as a function of the $\theta$. (b) Energy difference between FM and AFM phases (red line) and the Curie temperature (black line) under the biaxial strain for V$_2$AsP monolayer
\label{fig:64-E-b}}
\end{figure}

\section{CONCLUSION}

In summary, we have investigated the stability, electronic and magnetic properties of Janus V$_2$AsP monolayer by using first-principles calculations. The phonon spectrum calculations indicate that it is dynamically stable. The Janus V$_2$AsP monolayer with 100\% spin polarization is an FM half-metal. The half-metallic band gap is 0.38 eV and the spin gap for the semiconducting channel is 1.34 eV. Moreover, the MAE with an in-plane easy magnetization direction is 177.39 $\mu$eV. Monte Carlo simulations based on the Heisenberg model evaluate the \emph{T}$_c$ of 83 K for the V$_2$AsP monolayer, which is higher than recently reported \emph{T}$_c$ of CrI$_3$ (45 K) and Cr$_2$Ge$_2$Te$_6$ (30 K). A tensile strain of 4.9\% is applied to Janus V$_2$AsP monolayer, then a phase transition occurs from FM to AFM state. The \emph{T}$_c$ of Janus V$_2$AsP monolayer can be elevated to 130 K by a compressive strain of -5\%. The Janus V$_2$AsP monolayer might promote the development of spintronic nanoscale devices.

\begin{acknowledgments}
This work was supported by the Natural Science Foundation of Hebei Province (No. A2019203507 and B2021203030). The authors thank the High Performance Computing Center of Yanshan University.

\end{acknowledgments}


\bibliography{apssamp}

\end{document}